\documentclass[final,3p,twocolumn]{elsarticle}
\makeatletter
\def\ps@pprintTitle{%
  \let\@oddhead\@empty
  \let\@evenhead\@empty
  \let\@oddfoot\@empty
  \let\@evenfoot\@oddfoot
}
\makeatother
\usepackage{lineno,hyperref}
	\hypersetup{colorlinks=true,pdfborder=0 0 0}
\usepackage{subcaption}
\usepackage{amsmath}
\usepackage{amssymb}
\usepackage{gensymb}
\usepackage{cleveref}
\usepackage[group-separator={,}]{siunitx}
\usepackage{booktabs}
\usepackage[dvipsnames]{xcolor}
\usepackage{multirow}
\usepackage{float}
\modulolinenumbers[5]
\usepackage{graphicx}
\usepackage{tabularx,makecell}
\graphicspath{ {./figures/} }

\journal{Journal}

\begin{document}

\begin{frontmatter}

\title{Hearing the forest for the trees: \\ machine learning and topological acoustics for remote sensing with seismic noise}

\address[newfos]{New Frontiers of Sound Science and Technology Center, University of Arizona, Tucson, AZ 85721, USA}
\address[az-mse]{Department of Materials Science and Engineering, University of Arizona, Tucson, AZ 85721, USA}
\address[az-phys]{Department of Physics, University of Arizona, Tucson, AZ 85721, USA}
\address[az-geo]{Department of Geosciences, University of Arizona, Tucson, AZ 85721, USA}
\address[az-am]{Graduate Interdisciplinary Program in Applied Mathematics, University of Arizona, Tucson, AZ 85721, USA}
\address[ak]{-EWHALE Lab-, Department of Biology and Wildlife, Institute of Arctic Biology, University of Alaska, Fairbanks, AK 99775, USA}

\author[newfos,az-mse]{Jiayang Wang}
\author[newfos,az-phys]{I-Tzu Huang}
\author[newfos,az-geo]{Bingxu Luo}
\author[newfos,az-geo]{Susan L. Beck}
\author[newfos,ak]{Falk Huettmann}
\author[newfos]{Skyler DeVaughn}
\author[az-am]{Benjamin Stilin}
\author[newfos]{Keith Runge}
\author[newfos,az-mse]{Pierre Deymier}
\author[newfos,az-mse,az-am]{Marat I. Latypov\corref{cor}}
\cortext[cor]{corresponding author}
\ead{latmarat@arizona.edu}

\begin{abstract}

Monitoring remote forests is a global challenge central to climate mitigation and biodiversity conservation, yet satellite observations are frequently limited by weather, dense canopies, and solar dependency. Here we show that passive seismic sensing offers a persistent, all-weather alternative for autonomous ecosystem monitoring by capturing characteristic learnable signatures of trees within the ambient wavefield. Using seismic data from Alaska, we demonstrate that cross-correlations between stations provide a physical basis for forest detection by approximating the empirical Green's function of the medium. Supervised machine learning models applied to these data achieve a classification accuracy of \SI{86}{\percent}, identifying key discriminating frequencies (\SI{35}{} to \SI{60}{\hertz}) consistent with known forest--wave interactions. A topological acoustics analysis of the geometric phase change independently confirms the physical origin of these data-driven classifications. Together, these results provide the first demonstration that subtle forest--wave interactions manifest in ambient seismic noise and can be harnessed as a scalable tool for continuous vegetation monitoring, offering a robust solution for tracking environmental change challenging regions.

\end{abstract}

\begin{keyword}
Seismology \sep Remote sensing \sep Machine learning \sep Spectral analysis \sep Topological acoustics
\end{keyword}
\end{frontmatter}

\section*{Introduction}

Large-scale monitoring of environmental changes is critical for understanding and mitigating the accelerating impacts of climate change and deforestation. The global conservation efforts are often hindered by challenges in monitoring and data "blindness" in remote regions \cite{hansen2013high,zhu2021assessing}. Satellite remote sensing (radar, optical) is indispensable but its temporal consistency is often compromised by cloud cover, dense canopies, and lack of solar illumination (at night or due to weather conditions)  \cite{asner2001cloud,lucas2010evaluation}. These constraints create critical data gaps for forest ecosystems that are central to global carbon and water cycles \cite{bonan2008forests,pan2011large} yet face increasing pressures from cutting and other climate-driven disturbances \cite{chapin2011earth}. To address these challenges, seismic remote sensing is emerging as a complementary resilient modality \cite{carver2024polarization, colombi2016forests, chinmayi2024advancements}. By leveraging the propagation of elastic waves through the subsurface, seismic networks provide an all-weather autonomous monitoring framework capable of sensing forest dynamics \cite{abdalzaher2024emerging}.

\begin{figure*}[h]
  \centering
  \includegraphics[width=0.9\linewidth]{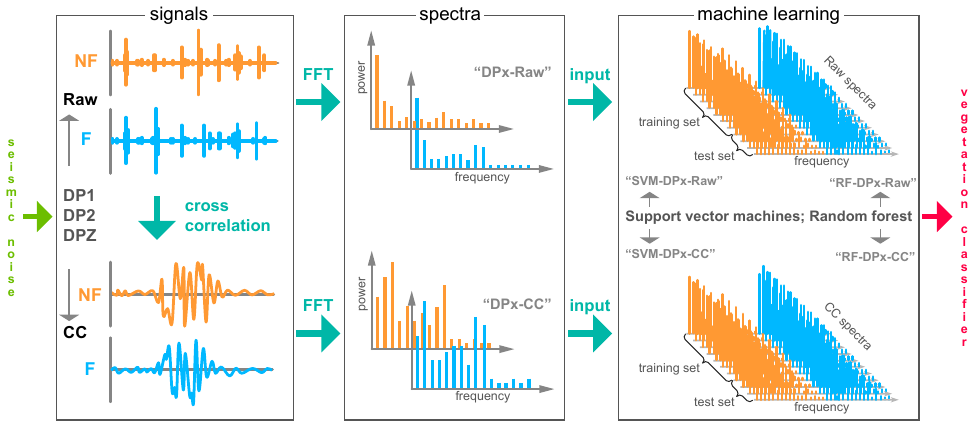}
  \caption{Graphical overview of the two machine learning workflows employed in this study based on the raw seismic data ("Raw") and cross-correlations (CC). DPx (x$\in\{1, 2, Z\}$) are the signal components considered in the study, multiple machine learning models considered are labeled as algorithm-component-input type (e.g., SVM-DPZ-CC).}
  \label{fig:workflow}
\end{figure*}

The utility of seismic sensing relies on the physical interaction between elastic waves and natural media. Recent research suggests that forests can exhibit behavior of natural seismic metamaterials owing to their heterogeneous structure, in which individual trees act as vertical resonators \cite{kadic2013metamaterials, brule2020emergence} that collectively modulate and attenuate surface (Rayleigh, Love) waves \cite{zhang2024attenuation}. The spatial arrangement and physical properties of the trees can give rise to such phenomena as band gaps and energy trapping that are typical of engineered metamaterials \cite{craster2012acoustic,colombi2016forests}.

Theoretical and numerical investigations have established that the metamaterial-like behavior of forests arises from specific tree--wave interactions. Trees act as subwavelength resonators that couple with the vertical component of Rayleigh waves and thus fundamentally alter their propagation \cite{colombi2016forests}. When arranged with a height gradient, trees form metawedges that enable mode conversion and wave localization, which creates multiple attenuation bands between \SI{10}{} and \SI{130}{\hertz} through cumulative resonances \cite{colombi2016seismic}. These behaviors are attributed to hybridization of Love and Rayleigh waves with tree resonances, resulting in subwavelength band gaps and energy trapping \cite{maurel2018conversion, muhammad2020forest, muhammad2021natural}. While efficiency of this coupling depends on local site conditions \cite{he2023forest}, the geometric parameters of the forest (tree height, trunk radius, and spatial density) remain the key controls for the frequency and width of these band gaps \cite{liu2019trees, al2024efficient}. Although these mechanisms have been experimentally validated \cite{brule2014experiments, colombi2016forests, roux2018toward} and explored for vibration mitigation \cite{pu2018surface, zhao2022theoretical, hao2024numerical} and infrastructure protection \cite{ouakka2024forests}, their potential as a diagnostic tool for ecological monitoring remains largely untapped. 

In this context, we hypothesize that forests introduce distinct and detectable signatures into ambient seismic noise. We furhter hypothesize that, if such signatures exist, they can be extracted from seismic data to discriminate forested from non-forested environments \textit{without the need for active seismic sources}.

\begin{figure}[h]
  \centering
  \includegraphics[width=\linewidth]{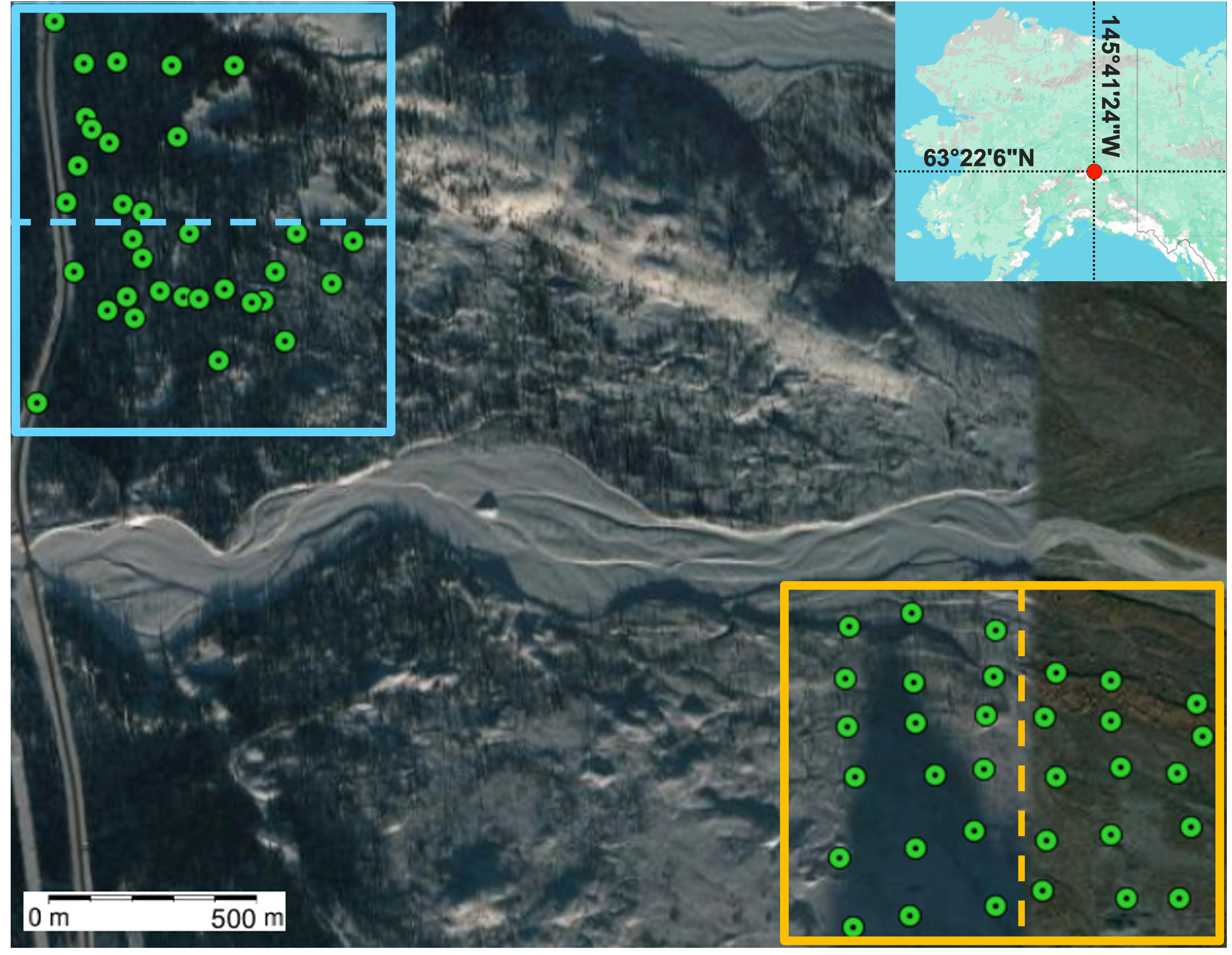}
  \caption{Geographic location of sensors in Alaska near Denali Fault and Richardson Highway intersection. The green dots represent the seismic stations whose signals were used. The blue and orange boxes indicate the forested and non-forested regions, respectively. The dashed lines show the split of the sensors within both regions for intra-class control analysis.}
  \label{fig:sensor_location}
\end{figure}

In this study, we develop and evaluate machine learning models to test these hypotheses by classifying forested vs.\ non-forested regions (see \Cref{fig:workflow} and Methods). Our approach is grounded in the principle that cross-correlations between signals from pairs of passive seismic sensors approximate the empirical Green’s function of the medium \cite{derode2003green,lobkis2001emergence} and thus encode its structural response. Models using spectra of these cross-correlations trained on a seismic dataset from Alaska \cite{allam2016} (\Cref{fig:sensor_location}) achieve an overall classification accuracy of \SI{86.2}{\percent}. %Models trained directly on spectra of raw signals performed worse, confirming the value of Green’s function representations.

To provide a physical validation for these data-driven results, we apply a topological acoustics approach to the geometric analysis of the cross-correlated wavefield \cite{deymier2017sound, alu2025bright}. In this formulation, seismic stations serve as discrete spatial probes of the medium, while the ensemble of pairwise empirical Green’s functions describes the geometry of wave propagation through the medium. Using this representation, we compute a geometric phase change ($\Delta\eta$) \cite{deymier2017sound, alu2025bright, Luo2025Geometric} across station pairs. The resulting $\Delta\eta$ spectra exhibit clear frequency-dependent differences between forested and non-forested regions in the frequency bands identified as most critical by machine learning classifiers. 

Together, these analyses confirm our hypotheses and demonstrate, for the first time (to our knowledge), that theoretically predicted forest--wave coupling can be detected passively in ambient seismic noise and exploited for practical classification in a real-world setting. This work establishes a vital bridge between the fundamental physics of seismic metamaterials and their application in sustainability-focused remote sensing with scalable potential for ecosystems where traditional satellite observations are limited. 

\section*{Results}

\subsection*{Approximating the Green's function from ambient seismic noise}

\begin{figure}[t]
  \centering
  \includegraphics[width=\linewidth]{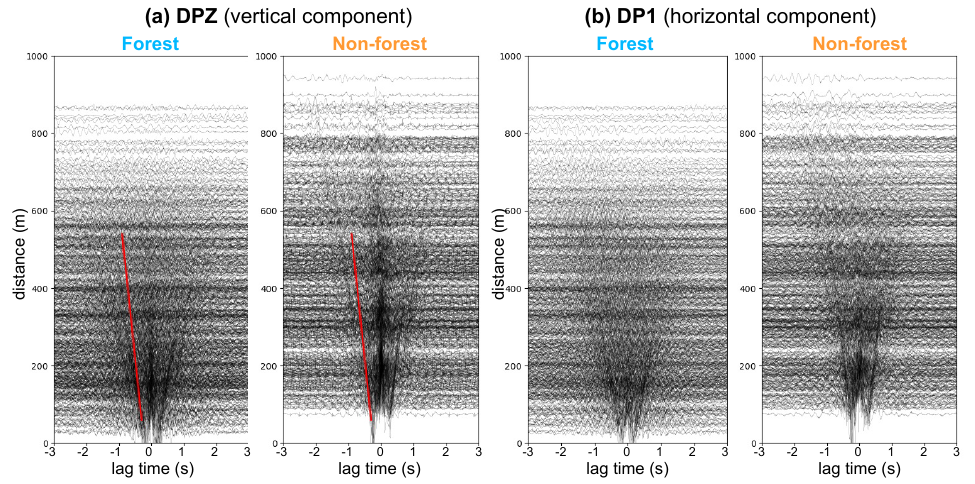}
  \caption{Cross-correlation record sections of (a) DPZ and (b) DP1 components of the signals from forested and non-forested regions obtained with a bandpass filter from \SI{10}{} to \SI{100}{\hertz}. Red lines indicate the approximate arrivals of primary phases with a velocity of \SI{700}{\meter\per\second}.}
  \label{fig:CC_distance_DPZ}
\end{figure}

The premise of our approach is that cross-correlations of ambient seismic signals approximate the empirical Green's function of the medium between recording stations. We evaluate this approximation using cross-correlation record sections \cite{snieder2008extracting, wapenaar2008unified}, which illustrate how well the correlations capture the wavefield between sensors. \Cref{fig:CC_distance_DPZ} shows such record sections of the cross-correlations calculated for the vertical (DPZ) and horizontal (DP1) components in both forested and non-forested regions. Each section displays 528 pair-wise cross-correlations vertically ordered by the inter-station distance. We focus on the central (zero-lag) peaks, which represent the correlation maximum corresponding to simultaneous signal arrival, and on the oscillations around zero lag (phase splitting), which reflect coherent wave arrivals from opposite directions. The progressive broadening of the zero-lag peaks with increasing distance (red lines in \Cref{fig:CC_distance_DPZ}) and the emergence of symmetric phase-splitting patterns indicate a well-diffused ambient noise field, which enables convergence towards an empirical Green’s function \cite{snieder2008extracting, campillo2008long}. Similar systematic broadening is observed for the second horizontal component (DP2, see Supplementary). These results confirm sufficient wavefield diffusivity for reliable estimation of the empirical Green's function. Consequently, the cross-correlations capture information about seismic wave propagation through the underlying medium, which provides a physical basis for machine learning presented in the following sections. 

\subsection*{Classification of forested and non-forested areas with machine learning}

\begin{figure*}[h]
  \centering
  \includegraphics[width=0.8\linewidth]{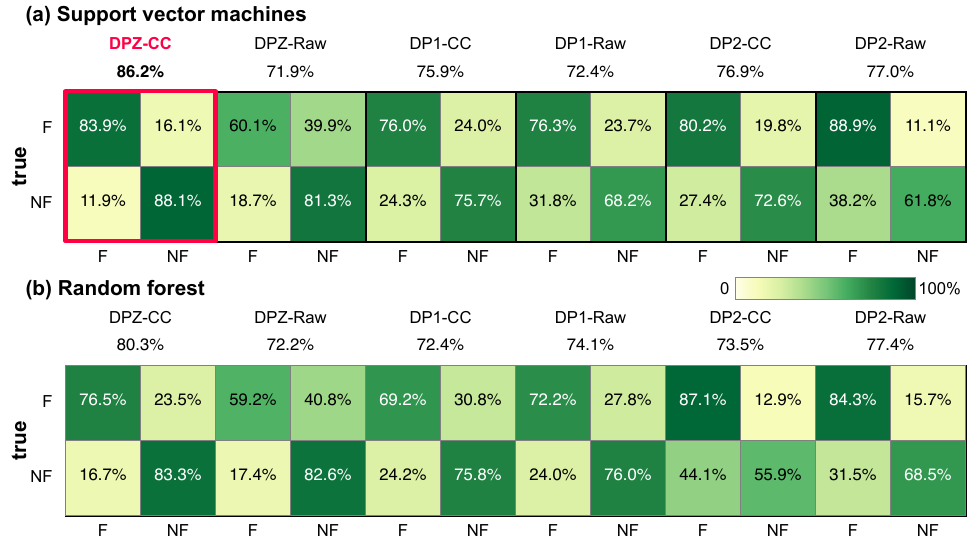}
  \caption{Confusion matrices of the (a) SVM and (b) RF models based on different input signals. Percentages (and the colors) in the cells represent true positive (TP) rate, false negative rate, true negative (TN) rate, and false positive rate (in clockwise order in every $2\times2$ confusion matrix). F indicates forest, NF non-forest, CC cross-correlation. The percentage values under the input signal labels represents overall classification accuracy defined as $(\text{TP}+\text{TN})/n$, where $n$ is the total number of samples in the given test set (\Cref{tab:dataset}).}
  \label{fig:confusion_matrix}
\end{figure*}

\begin{figure}[h]
  \centering
  \includegraphics[width=\linewidth]{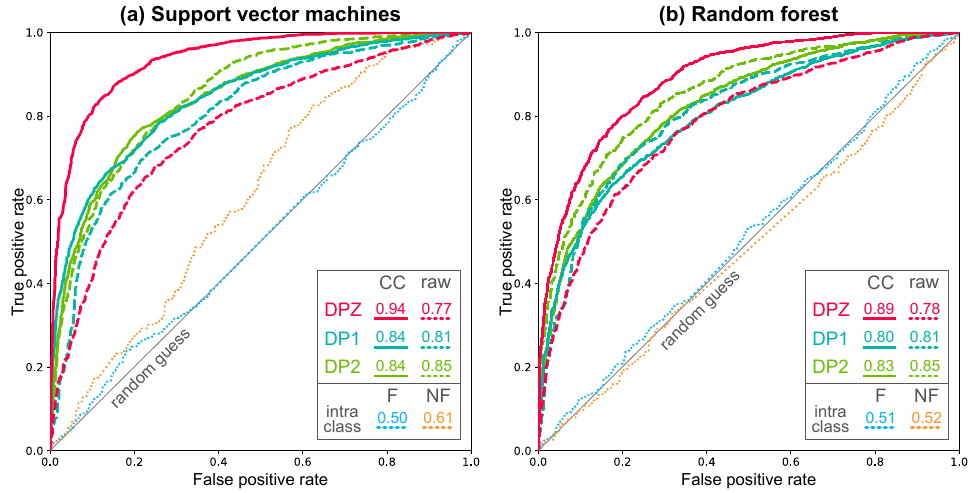}
  \caption{ROC curves obtained for six input signals (a) using SVM, (b) using RF. The solid lines and dashed lines represent cross-correlation (CC) and raw signals (raw), respectively. Dotted lines represent ROC curves from intra-class analysis within forested (F) and non-forested (NF) regions (obtained with SVM of the DPZ cross-correlations).}
  \label{fig:ROC_curve}
\end{figure}

Having established approximation of the empirical Green’s function with the cross-correlations, we next examine whether they contain signatures sufficient to distinguish forested from non-forested environments. To this end, we trained supervised machine learning models using two complementary algorithms, support vector machines (SVM) and random forests (RF), and evaluated their performance on two types of inputs: cross-correlations as well as the underlying raw signals as a baseline. Upon training, the classification models exhibit strong discriminatory power as evident from the confusion matrices (\Cref{fig:confusion_matrix}), receiving operating characteristic (ROC) curves (\Cref{fig:ROC_curve}), and corresponding areas under the ROC curves (AUC). Among tested model and input combinations (see Methods), the SVM trained on cross-correlations of the vertical (DPZ) component as input achieved the best results: a true positive rate of \SI{88.1}{\percent}, a true negative rate of \SI{83.9}{\percent}, and an overall accuracy of \SI{86.2}{\percent} on the test set unseen during training. In absolute terms, it correctly classified \SI{1418}{} of \SI{1690}{} forested samples and \SI{1860}{} of \SI{2112}{} non-forested samples. These results correspond to a precision of 0.85, recall of 0.84, and F1-score of 0.84, indicating a balanced accuracy across both classes.  Models trained on cross-correlations generally outperformed those using raw seismic signals, confirming that the empirical Green’s function provides more discriminative features of the underlying medium (see \Cref{fig:confusion_matrix}). 

Analysis of ROC curves further supports these findings: all models performed well above random classification ($\text{AUC}>0.77$), with the DPZ-based SVM approaching near-perfect separability ($\text{AUC}=0.94$,  \Cref{fig:ROC_curve}). Collectively, these results demonstrate that cross-correlations of the ambient noise encode physically meaningful signatures of trees that can be systematically learned with algorithms. To verify that the models learned meaningful distinctions rather than spurious patterns in any pair of datasets, we performed an intra-class control test. Using the same SVM and RF architectures re-trained on subsets of data within each class (forested or non-forested), we found that classification performance dropped to levels comparable to random guess ($\text{AUC} \in [0.5,0.6]$; \Cref{fig:ROC_curve}). These results confirm that the models do not artificially separate signals within a single class and that the observed discrimination indeed reflects intrinsic differences between forested and non-forested environments.

To interpret what the classifiers have learned, we analyzed feature importance and partial dependence plots across spectral frequencies (\Cref{fig:feature_importance}). Focusing on the DPZ cross-correlations, which yielded the highest classification accuracy, both SVM and RF models consistently identified four dominant frequencies as the most influential for distinguishing forested from non-forested areas: \SI{46.83}{\hertz}, \SI{35.16}{\hertz}, \SI{58.5}{\hertz}, and \SI{11.83}{\hertz}. Partial dependence analysis further confirms the importance of these frequencies ($\omega_1$ through $\omega_4$): variations in their amplitudes have the highest impact on the probability of predicting the forest class with steep slopes compared to lower-ranked frequencies ($\omega_5$ through $\omega_8$). The contrast is particularly pronounced for the RF model, whose step-like curves reflect the discretized nature of its decision trees, whereas the smoother curves of the SVM model arise from its continuous decision boundaries \cite{friedman2009elements, molnar2020interpretable}.

\subsection*{Distinction of forested and non-forested areas with topological acoustics}

\begin{figure}[h]
  \centering
  \includegraphics[width=\linewidth]{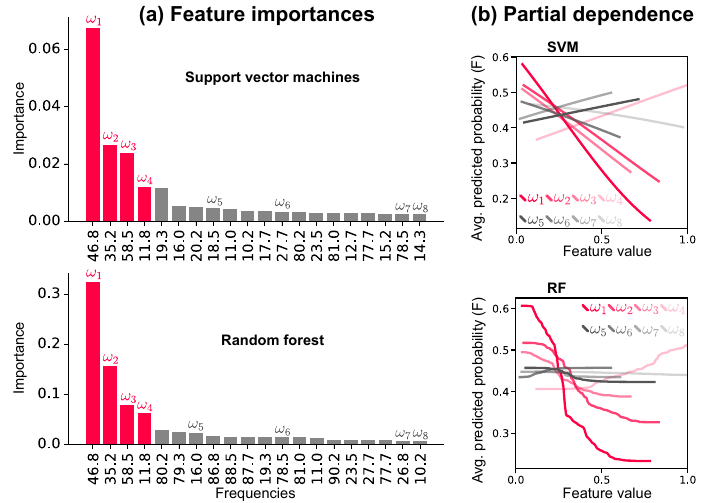}
  \caption{(a) Feature importances and (b) partial dependence plots for the SVM and RF models using cross-correlations of the DPZ signal as input.}
  \label{fig:feature_importance}
\end{figure}

The frequency-dependent contrasts revealed by machine learning models suggest that the forest--wave coupling alters the underlying structure of the seismic wavefield. To investigate these effects from a topological perspective, we compute the average geometric phase change, $\Delta \eta$, for the forested and non-forested regions as a function of frequency (\SI{10}{} to \SI{100}{\hertz}). The average $\Delta \eta$ integrates over all possible state vectors within each region (see \Cref{fig:average_phase_change}a and Methods), providing a direct measure of the collective wavefield. \Cref{fig:average_phase_change} shows $\Delta \eta(\omega)$ for three dimensions of the state vector ($N=4, 7, 10$). The overall amplitude of $\Delta \eta$ increases with $N$ and stabilizes beyond $N=7$ indicating convergence with respect to the state vector dimension. Across all three dimensions, the forested and non-forested regions exhibit distinct phase behavior: both show elevated $\Delta \eta$ at low frequencies (below \SI{42}{\hertz}), but clear divergences appear in the bands around \SI{10}{\hertz} and from \SI{18}{} to \SI{42}{\hertz}. These are precisely the frequency ranges that dominated the feature importances from machine learning (\Cref{fig:feature_importance}), which confirms that the presence of trees is encoded in the topology of the wavefield.

\section*{Discussion}

The high classification accuracy of machine-learning models, together with distinct geometric phase changes observed between forested and non-forested regions, demonstrates that trees leave measurable signatures in the seismic wavefield. We confirmed this finding using both raw seismic signals and inter-station cross-correlations as inputs for machine-learning algorithms as well as through geometric phase analysis across multiple state vector dimensions. In all cases, signals from forested and non-forested areas were separable both quantitatively through machine learning and qualitatively through geometric phase change. Several lines of evidence reinforce that this distinction reflects systematic, physically grounded signatures of the forest--wave interaction: (i) cross-correlations converged to an empirical Green’s function; (ii) intra-class analysis yields near-random performance, ruling out spurious separation; (iii) the frequency bands most important for classification correspond to those where theoretical forest effects are strongest; (iv) DPZ cross-correlations associated with Rayleigh waves achieve the highest accuracy; and (v) the average geometric phase, $\Delta\eta$, exhibits consistent differences between forested and non-forested regions within the same frequency range at all studied state vector dimensions.

\begin{figure*}[h]
  \centering
  \includegraphics[width=0.9\linewidth]{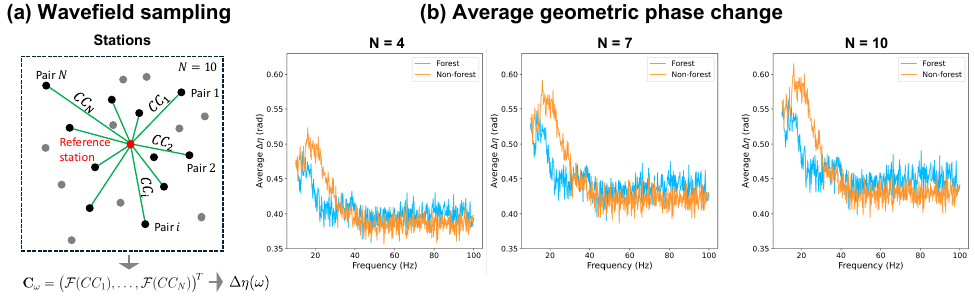}
  \caption{Geometric phase analysis: (a) schematic of state vector construction for given a group of sensors (red symbol for reference station, black: stations paired for cross-correlations (CC), gray: stations unused in the current combination); (b) averaged geometric phase change, $\Delta \eta(\omega)$ for the forested and non-forested regions at different state vector dimensionalities ($N=4, 7, 10$). Simplified equations are provided, see the Supplementary for complete expressions.}
  \label{fig:average_phase_change}
\end{figure*}

First, the premise of this work is that the empirical Green's function is encoded in cross-correlations of seismic noise. Our cross-correlation record sections (\Cref{fig:CC_distance_DPZ}) confirm this premise for the seismic signals considered here, and the superior classification accuracy obtained from cross-correlations relative to raw signals (\Cref{fig:confusion_matrix,fig:ROC_curve}) further supports it. 

Second, the intra-class analysis showed that distinguishing signals from actual forested and non-forested regions yields much higher accuracy than distinguishing signals within either region alone. This result confirms that the models capture macroscale structural differences linked to trees rather than arbitrary environmental variance.

Third, the frequency bands identified as most influential for classification (\Cref{fig:feature_importance}) agree with frequencies known to characterize forest--wave interactions. Specifically, key frequencies identified in our study (\SI{47}{}, \SI{35}{}, and \SI{59}{\hertz}) overlap with experimentally observed forest-induced band gaps in the vertical component (\SI{45}{} to \SI{60}{\hertz}) \cite{roux2018toward} and numerically predicted attenuation ranges (\SI{30}{} to \SI{48}{\hertz} and \SI{57}{} to \SI{65}{\hertz}) associated with forest metawedges \cite{al2024efficient}.

Fourth, cross-correlations of the vertical DPZ component lead to the highest classification accuracy (\SI{86.2}{\percent}), outperforming the models based on the horizontal components (DP1, DP2). The DPZ-based models also showed the largest improvement when using cross-correlations instead of raw signals (+\SI{14.3}{\percent} for SVM, \Cref{fig:confusion_matrix}a). This result is physically significant because cross-correlations of the vertical component are strongly associated with Rayleigh waves \cite{savage2013ambient, takagi2014separating}, which are particularly sensitive to attenuation by forests and their characteristics (e.g., tree height and spacing) \cite{muhammad2021natural, he2023forest, colombi2016forests}. 

Finally, the average geometric phase change, $\Delta \eta$, reveals clear distinction between forested and non-forested areas within specific frequency bands. These bands align with the important frequencies identified by the machine-learning models and previously reported interactions such as forest-induced band gaps \cite{craster2012acoustic,colombi2016forests}. This agreement demonstrates that the geometric phase $\Delta \eta$, as a novel acoustic signal processing metric, captures essential information about forest--waves interactions encoded in the ambient seismic field. The reduced dispersion of the geometric phase in forested regions found in this study (\Cref{fig:average_phase_change}) is likely to reflect resonance of trees within a narrow frequency band (e.g., due to similar height), as has been numerically studied \cite{lata2022topological}.

The strong classification performance and consistent geometric phase differences demonstrate the viability of using ambient seismic noise as a remote sensing tool for forest ecosystems. Unlike high-intensity but rare seismic events (e.g., earthquakes), ambient noise is continuous and ubiquitous, which renders it a practical, scalable, and resilient resource for monitoring natural environments. %By linking learnable forest signatures with the fundamental physics of forest--wave interaction, this work establishes a new modality for ecosystem surveillance. 

Advancing this framework from proof-of-concept to practical application will involve further development and validation. While this study focused on a specific region in Alaska as a proof of concept, the physically grounded nature of the features (empirical Green's functions) is fundamentally generalizable to other geographic, ecological, and seasonal settings. Since the key frequencies identified here are attributed to tree geometry \cite{liu2019trees, al2024efficient}, we expect that forests with different structural properties will exhibit shifted but conceptually analogous and learnable spectral signatures. Seasonal variations and changes in forest health are expected to introduce similar shifts. Further seismic measurements can validate these expectations. To handle the complexity of large and diverse datasets across the globe as well as generalization to new settings, future work could employ more sophisticated machine learning methods  (e.g., convolutional neural networks or ensembles of classifiers). Finally, complementary numerical simulations could help map causal relationships between forest structures and seismic features for high-precision sensing of biomass estimation, health monitoring, and disturbance detection beyond binary classification tackled here.

\section*{Conclusions}

This study tested the hypothesis that ambient seismic noise carries measurable signatures of trees in forested areas that can be systematically decoded using supervised machine-learning algorithms and distinguished with topological acoustics. Applying these approaches to continuous seismic data from Alaska led to the following conclusions: 

\begin{enumerate}

\item Forested and non-forested areas were successfully distinguished using machine learning with overall accuracies ranging from \SI{71.9}{\percent} to \SI{86.2}{\percent} depending on the signal component (horizontal vs.\ vertical) and input type (raw vs.\ cross-correlation). %and, to a lesser extent, the choice of the learning algorithm (support vector machines vs.\ random forests). 

\item Cross-correlations of the vertical (DPZ) component achieved the most robust classification results (AUC of 0.94, F1 score of 0.84), which stems from the effective retrieval of the empirical Green's function and the sensitivity of Rayleigh waves to vegetation-induced scattering.

\item The most discriminative frequencies (between 30 and \SI{60}{\hertz}) align with the forest-induced band gaps and attenuation ranges reported in experimental and numerical studies of forest--wave interactions, supporting the physical basis of the machine learning results. 

\item The average geometric phase change, $\Delta\eta$, provides an independent physical metric for forest detection; the differences in the spectra in forested vs.\ non-forested regions occur within the same critical frequency ranges identified by the classifiers and thus offers a topological confirmation of the data-driven results. 

\end{enumerate}

The convergence of the data-driven and topological acoustics results  suggests that subtle forest--wave interactions can be harnessed for scalable environmental sensing. This work builds the foundation for using passive seismic monitoring as a resilient, all-weather, and scalable modality for tracking forest dynamics and as a vital redundancy to traditional satellite methods in monitoring the Earth's changing ecosystems. 

\section*{Methods}
\label{sec:methods}

\subsection*{9M Dataset overview}

This study used open seismic waveform data from the 9M network operated by Allam et al.\ \cite{allam2016} in Alaska between April and May in 2016. The array included about 200 nodal sensors deployed near the Denali Fault and Canwell Glacier (\Cref{fig:sensor_location}). Each sensor recorded three components: "DP1" (horizontal, North--South), "DP2" (horizontal, East--West), and "DPZ" (vertical) at a sampling rate of \SI{2000}{\hertz}. Metadata such as time period, sensor locations, and component types were used for sensor selection and labeling.

For machine learning, two groups of sensors were selected from areas with distinct vegetation densities (\Cref{fig:sensor_location}): an area densely populated with trees (``forest'' class)  and a sparsely vegetated area (``non-forest'' class). In total, 66 sensors (33 per class) were selected. The corresponding waveform data were downloaded in the \texttt{mseed} format from the EarthScope Data Center (IRISDMC) \cite{allam2016} with each file containing a continuous record over \SI{24}{\hour}. Although the full 9M dataset covers nearly one month, the recordings from the selected sensors were partially incomplete ranging from 15 to 20 days, depending on the component. Only the days with complete data were used for machine learning.

\subsection*{Data processing}
%\label{sec:data}

The raw waveform data were processed through two workflows (\Cref{fig:workflow}): one using raw signals directly and another based on their cross-correlations, which approximate the empirical Green's function between sensors. Both workflows resulted in spectra in the frequency domain that served as predictor variables for machine-learning classification. 

\emph{Processing of the raw signals}. The first workflow involved (i) segmenting raw signals into 10-min intervals, (ii) selecting time windows of interest, and (iii) transforming the segments into the frequency domain using fast Fourier transform (FFT). The high sampling rate of the 9M dataset (\SI{2000}{\hertz}) produces about nine million data points per \SI{24}{\hour} resulting in an FFT frequency resolution of $5.79\times 10^{-9}$ \SI{}{\hertz}. To reduce the computational cost, each \SI{24}{\hour} recording was divided into 10-min segments covering time windows from 6:00 to 8:00 (six intervals) and from 22:00 to 24:00 (six intervals, see Supplementary). These time windows were selected to minimize anthropogenic noise (e.g., road traffic). Sample counts for all three components (DP1, DP2, DPZ) are summarized in \Cref{tab:dataset}. Each time-domain sample was transformed to the frequency domain using FFT over the range of \SI{10}{} to \SI{100}{\hertz}. This range captures frequencies most relevant to forest--wave interactions (below \SI{100}{\hertz} \cite{roux2018toward, colombi2016forests, lott2020evidence, he2023forest, maurel2018conversion, muhammad2021natural, baker1997measurements, al2024efficient}) while avoiding low frequencies 9M sensors have limited sensitivity (below \SI{5}{\hertz}). One hundred evenly spaced frequencies within this range were selected as features for machine learning with spectral power as the feature value.

\emph{Calculation of cross-correlation signals}. The second workflow computed cross-correlations of ambient signals as an empirical representation of the Green's function between sensors based on the time-reversal symmetry and wavefield reciprocity \cite{derode2003green}. Cross-correlations were computed using the \texttt{MSNoise} Python package \cite{lecocq2014msnoise}. Raw signals were divided into \SI{1800}{\second} intervals with \SI{50}{\percent} overlap before processing. Combining 33 sensors into 528 unique pairs results in 528 cross-correlations per class per day and thus thousands of cross-correlation samples per class over $>2$ weeks in total (\Cref{tab:dataset}). %For example, 15 days of DP1 recordings provides $528 \cdot 16 = 8448$ cross-correlation samples for the forested region; the counts for all three components and both regions are summarized in \Cref{tab:dataset}. 
Each cross-correlation ``signal'' was treated similar to a raw waveform: FFT was used to obtain power spectra followed by adoption of 100 evenly spaced frequencies as features for classification, which ensured the consistency of the two workflows and allowed direct comparison of the resulting models. 

\begin{table*}[h]
    \centering
    \begin{tabular}{ccccc}
        \toprule
        Component & Type of Signal & Forest & Non-Forest & Total \\
        \midrule
        \multirow{2}{*}{DP1} & Raw & \num{6336} & \num{5940} & \num{12276} \\
                             & CC  & \num{8448} & \num{7920} & \num{16368} \\
        \multirow{2}{*}{DP2} & Raw & \num{7128} & \num{5544} & \num{12672} \\
                             & CC  & \num{9504} & \num{7392} & \num{16896} \\
        \multirow{2}{*}{DPZ} & Raw & \num{6336} & \num{7920} & \num{14256} \\
                             & CC  & \num{8448} & \num{10560} &\num{ 19008} \\
        \bottomrule
    \end{tabular}
    \caption{Summary of the six datasets for machine learning. CC stands for cross-correlation.}
    \label{tab:dataset}
\end{table*}

\subsection*{Machine learning classification using SVM and RF}

To systematically distinguish seismic signals from forested and non-forested regions, we employed two widely used machine learning algorithms: support vector machines (SVM) \cite{vapnik1995support, scholkopf1997kernel} and random forests (RF) \cite{breiman2001random}. %These models were selected for their complementary strengths: the ability of SVM to learn complex, high-dimensional decision boundaries \cite{vapnik1995support} and the robustness of RF against noise and overfitting through ensemble averaging \cite{biau2012analysis}. 
Both SVM and RF models were trained and evaluated using the datasets described in the Data Processing section. To ensure models captured the relative distribution of spectral features rather than absolute amplitude, the spectral magnitudes of each sample were scaled to the range $[0,1]$. The scaled data and the corresponding forest and non-forest labels were randomly divided into \SI{80}{\percent} for training and \SI{20}{\percent} for testing with stratified sampling. %Stratified sampling was applied to preserve the original class ratio in both subsets. 
Model hyperparameters were optimized using $k$-fold cross-validation on the training data.  Model accuracy was evaluated with standard metrics, including overall accuracy, precision, recall, F1-score, receiver operating characteristic (ROC) curves, areas under the ROC curve (AUC), and confusion matrices (see Supplementary for definitions). Six datasets (three components times two input types) were used in total, each evaluated with SVM and RF classifiers, resulting in 12 trained and tested models. Details of Python implementation (with \texttt{scikit-learn} \cite{pedregosa2011scikit}), tuned hyperparameters, learning curves, intra-class control analysis, and theoretical background of the algorithms are provided in the Supplementary Materials. %\Cref{fig:workflow} summarizes the data processing and machine learning workflows. 

\subsection*{Geometric phase} 

We computed the average geometric phase change, $\Delta\eta(\omega)$, for the DPZ cross-correlation spectra as a novel metric for wavefield analysis \cite{deymier2017sound, Luo2025Geometric}. For each region, we constructed $N$-dimensional state vectors ($N=4, 7, 10$) from combinations of station pairs and computed the average $\Delta\eta(\omega)$ relative to a reference vector. Full computational details including sampling strategy and convergence testing are provided in the Supplementary.

\section*{Data availability}

The continuous seismic recordings (network code: 9M) \cite{allam2016} used in this study are open and available at the EarthSCope Data Management Center (IRISDMC, \url{http://service.iris.edu/fdsnws/dataselect/1/}). Ambient noise cross-correlations were computed using the open-source \texttt{MSNoise} package \cite{lecocq2014msnoise}. The data processing and machine learning were employed using Python with the aid of \texttt{scikit-learn} \cite{pedregosa2011scikit} and \texttt{matplotlib} \cite{hunter2007matplotlib} library. Codes used in this study is available at GitHub: \url{https://github.com/materials-informatics-az/forest-classification}.

\section*{Acknowledgements}

This research was supported by the New Frontiers of Sound (NewFoS) Science and Technology Center at the University of Arizona sponsored by U.S.\ National Science Foundation (Grant Number 2242925).

\section{Supplementary material}

\subsection{Data selection, Green's function, and cross-correlations}

\subsubsection{Data availability and time windows selected for raw signals}

The 9M dataset contains the following number of days with complete data: for the DP1 component, 16 full days for forest and 15 days for non-forest sensors; for DP2, there were 18 and 14 days, respectively; and for DPZ, 16 and 20 days. The raw 24-h signals from these complete days were segmented into 10-min intervals covering the time windows from 6:00 to 8:00 (six intervals) and from 22:00 to 24:00 (six intervals). The exact time windows used were the following: 6:00$\sim$6:10, 6:20$\sim$6:30, 6:40$\sim$6:50, 7:00$\sim$7:10, 7:20$\sim$7:30, 7:40$\sim$7:50, 22:00$\sim$22:10, 22:20$\sim$22:30, 22:40$\sim$22:50, 23:00 $\sim$23:10, 23:20$\sim$23:30, and 23:40$\sim$23:50. Each 10-min segment was treated as an independent sample, providing $16 \cdot 12 \cdot 33 = 6336$ samples labeled as ``forest'' and $15 \cdot 12 \cdot 33 = 5940$ ``non-forest'' samples for the DP1 component; thus resulting sample counts for other components are listed in Table 1 of the main text. 

\subsubsection{Green's function and cross-correlation record sections}

\emph{Green's function}. Empirical Green's functions were obtained by cross-correlating seismograms between all unique pairs of the sensors within the forested or non-forested regions. This approach is motivated by the theoretical framework of time-reversal symmetry and wavefield reciprocity, which suggests that the Green’s function between two passive sensors can be approximated by the cross-correlation of ambient seismic noise recorded at those locations \cite{derode2003green}. Formally, the cross-correlation function between two signals $s_A(t)$ and $s_B(t)$ recorded at sensors $A$ and $B$ is defined as

\begin{equation}
    C_{AB}(t) = \int s_A(\tau) \cdot s_B(\tau + t) \, d\tau.
\end{equation}

\noindent Assuming the energy is uniformly distributed across all natural frequencies, the expression reduces to the symmetrized Green’s function:

\begin{equation}
    C_{AB}(t) = A_0 h_{AB}(t),
\end{equation}

\noindent where $A_0$ is an amplitude scaling factor and $h_{AB}(t)$ is the Green's function between the passive sensors $A$ and $B$. This method enables passive estimation of inter-station impulse responses without requiring the knowledge of source location or timing, while reducing incoherent noise.

\begin{figure}[H]
  \centering
  \includegraphics[width=\linewidth]{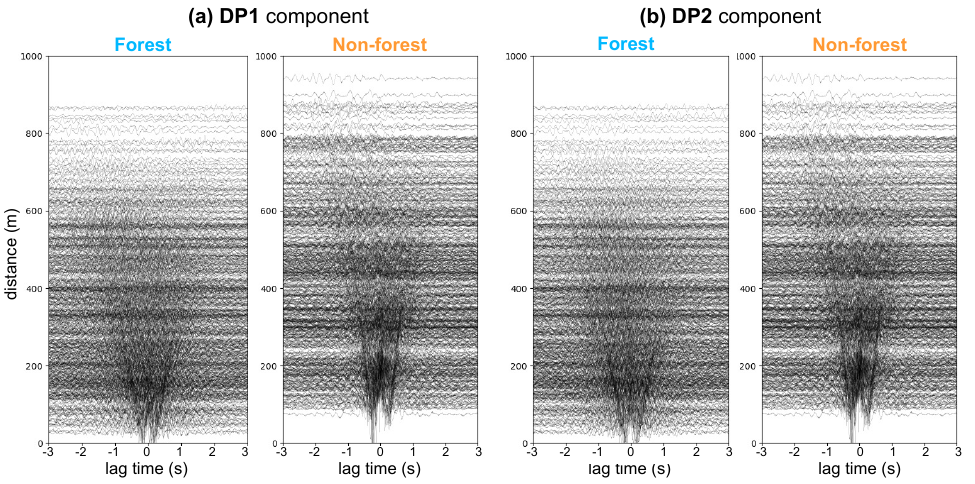}
  \caption{Cross-correlation record sections of (a) DP1 and (b) DP2 components of the signals from forested and non-forested regions obtained with a bandpass filter from \SI{10}{} to \SI{100}{\hertz}.}
  \label{fig:CC_distance_DP2}
\end{figure}

\emph{Cross-correlation record sections}. \Cref{fig:CC_distance_DP2} shows record sections of cross-correlations of the DP1 and DP2 components to demonstrate that the discussion given in the main text applies to all three components of the seismic signals considered in this study. 

\subsection{Machine learning models}

The two machine learning algorithms explored in this study were selected for their complementary strengths: the ability of support vector machines (SVM) to learn complex, high-dimensional decision boundaries \cite{vapnik1995support} and the robustness of random forests (RF) against noise and overfitting by ensemble averaging \cite{biau2012analysis}.

\subsubsection{Background}

\emph{Support vector machines} classifier was used as one of the machine learning algorithms used in this study because of its capability to handle high-dimensional classification problems with non-linear decision boundaries \cite{cortes1995support}. Training an SVM with a kernel $k(\hat{x_i},\hat{x_j})$ is formulated as an optimization problem of finding Lagrange multipliers $\alpha_i$ that maximize

\begin{equation}
    % \max_{\boldsymbol{\alpha}} 
    \sum_{i=1}^{n} \alpha_i 
    - \frac{1}{2} \sum_{i=1}^{n} \sum_{j=1}^{n} 
    \alpha_i \alpha_j y_i y_j k(\hat{x_i}, \hat{x_j})
\end{equation}

\noindent subject to the constraints

\begin{equation}
0 \leq \alpha_i \leq C, \quad \sum_{i=1}^{n} \alpha_i y_i = 0,
\end{equation}

\noindent where $\hat{x_i}$ and $y_i$ are the $i^\text{th}$ pair of the input vector and label, $C$ is the regularization parameter that controls the trade-off between maximizing the margin and minimizing classification errors \cite{cortes1995support}. In this study, we adopt SVM with an radial basis function kernel expressed as \cite{vapnik1995support}:

\begin{equation}
    k(\hat{x_i}, \hat{x_j}) = \exp(-\gamma \|\hat{x_i} - \hat{x_j}\|^2),
\end{equation}

\noindent where the kernel coefficient $\gamma > 0 $ controls the degree of nonlinearity of the hyperplane. 

\emph{Random forests} is the second representative classification algorithm used in this study. RF is an ensemble learning method that constructs decision trees and aggregates their predictions to achieve a high classification accuracy and reduce the risk of overfitting \cite{breiman2001random}. Decision trees are constructed by recursively splitting the target variable based on predictor variables from the parent node until a stopping criterion is met \cite{breiman2001random}. Each decision tree serves as a simple classifier contributing to the overall output, which is obtained by averaging the predictions of all individual trees. Key hyper-parameters of a typical RF model include: (i) number of trees (estimators), more trees generally improve accuracy but increase computation time; (ii) maximum tree depth, which controls the complexity of individual trees, with deeper trees capturing more patterns but risking overfitting; (iii) minimum number of samples required to split a node, where higher values may lead to underfitting; (iv) minimum number of samples required at a leaf node, which regularizes tree growth and prevents overly specific splits; and (v) maximum number of features, which balances the model diversity and predictive performance.

\subsubsection{Tuned hyper-parameters}

The hyper-parameters mentioned above were optimized in the two models with each of six inputs considered in this study using $k$-fold cross-validation on the corresponding training datasets. \Cref{tab:hyperparameters} lists the optimized values for all six input cases. 

\begin{table*}[h]
\centering
\renewcommand{\arraystretch}{1.2} % Increase row height
\begin{tabularx}{\textwidth}{l X X X X X X} %{l c c c c c c}

\toprule
\textbf{Param.\ \textbackslash{} Input} & \makecell{DP1 \\ Raw} & \makecell{DP1 \\ CC} & \makecell{DP2 \\ Raw} & \makecell{DP2 \\ CC} & \makecell{DPZ \\ Raw} & \makecell{DPZ \\ CC} \\

\midrule 

\multicolumn{7}{l}{\textbf{Support vector machines}} \\
\midrule 
C & 0.6 & 0.2 & 0.4 & 0.2 & 0.1 & 0.3 \\
$\gamma$ & 0.2 & 0.1 & 1 & 0.1 & 0.8 & 0.2 \\

\midrule

\multicolumn{7}{l}{\textbf{Random forest}} \\
\midrule 
n\_estimators & 400 & 500 & 300 & 500 & 300 & 400 \\
max\_depth & 15 & 15 & 6 & 15 & 5 & 14 \\
min\_samples\_split & 80 & 80 & 30 & 80 & 40 & 80 \\
min\_samples\_leaf & 30 & 90 & 150 & 50 & 160 & 80 \\
max\_features & 0.6 & 0.4 & 0.3 & 'log2' & 0.7 & 0.35 \\

\bottomrule
 
\end{tabularx}
\caption{Summary of hyper-parameters obtained for the SVM and RF models with different input signals.}
\label{tab:hyperparameters}
\end{table*}

\subsubsection{Accuracy metrics}

The machine learning models were evaluated and compared using confusion matrices, receiver operating characteristic (ROC) curves, learning curves, and the following metrics: areas under the ROC curves (AUC), precision, recall, and F1 score. The analyses and metrics described below and in the main text are adopted for the problem of binary classification pursued in this study. For notation, we consider forest as "positive" and non-forest as "negative" outcome of an observation or prediction. 

Confusion matrices presented in the main text (Fig.~4) display four fractions  ("rates") quantifying how the classifiers performed on each class: true positive rate (TPR), false negative rate (FNR), false positive rate (FPR), and true negative rate (TNR). These values are calculated and arranged in the matrix form as follows: 

\begin{equation}
\begin{pmatrix}
\text{TPR} & \text{FNR} \\[6pt]
\text{FPR} & \text{TNR}
\end{pmatrix}
=
\begin{pmatrix}
\dfrac{\mathrm{TP}}{\mathrm{TP}+\mathrm{FN}} & \dfrac{\mathrm{FN}}{\mathrm{TP}+\mathrm{FN}} \\[12pt]
\dfrac{\mathrm{FP}}{\mathrm{TN}+\mathrm{FP}} & \dfrac{\mathrm{TN}}{\mathrm{TN}+\mathrm{FP}}
\end{pmatrix},
\end{equation}

\noindent where TP are true positives, FN are false negatives, FP are false positives, and TN are true negatives. Note that the actual (true) values are represented by rows so that the rates in each row add up to \SI{100}{\percent}.

TPR is also known as the recall (or sensitivity), while TNR is often referred to as the negative recall (or specificity). In addition to values in the confusion matrix, two other common metrics relevant to classification problems are the precision defined as:

\begin{equation}
    \text{Precision} = \frac{\text{TP}}{\text{TP} + \text{FP}}
\end{equation}

\noindent and the F1-score, which represents the harmonic mean of precision and recall and combines both metrics into a single value:

\begin{equation}
\text{F1-score} = 2 \times \frac{\text{Precision} \times \text{Recall}}{\text{Precision} + \text{Recall}}.
\end{equation}

The overall classification accuracy reported in the main text and Fig.~4 is defined as the number of correct predictions for both negative and positive samples divided by the total number of samples in the dataset of interest:

\begin{equation}
    \text{Accuracy} = \frac{\text{TP+TN}}{\text{TP}+\text{TN}+\text{FP}+\text{FN}}.\\
\end{equation}

ROC curves visualize the fraction of true positives (TPR) versus the fraction of false positives (FPR) at various classification thresholds \cite{bradley1997use}. Each pair of FPR and TPR corresponds to a single classification threshold and thereby form one point on the ROC curve. The ROC curve of a perfect classifier goes straight up the vertical axis ($\text{TPR} = 1$) and then across the top ($\text{FPR} = 0$), forming a sharp corner. In contrast, random guess corresponds to a line formed by $\text{TPR}=\text{FPR}$ (\SI{45}{\degree} to the horizontal axis). Quantitatively, a ROC curve is described by the area under the curve (AUC): $\int_{0}^{1} \text{TPR(FPR)}d(\text{FPR})$. According to the character of ROC curves described above, AUC of 1 would describe perfect classification, while an AUC of 0.5 corresponds to random guessing.

\subsubsection{Intra-class control analysis}

In addition to standard evaluation using metrics above, intra-class control analysis was used to verify that the trained models captured genuine differences between forest and non-forest regions rather than artifacts from internal variations within each class. For this purpose, sensors within each region were subdivided into two spatially distinct subgroups (dashed lines in Fig.\ 2 of main text). In the forested area, the two subgroups contained 13 and 20 sensors, producing 78 and 190 unique cross-correlation pairs, respectively, and were labeled ``F1'' and ``F2''. Likewise, the non-forested region was subdivied into subgroups of 18 and 15 sensors, yielding 153 and 105 cross-correlation pairs, labeled ``NF1'' and ``NF2''.

\begin{figure*}[h]
  \centering
  \includegraphics[width=0.9\linewidth]{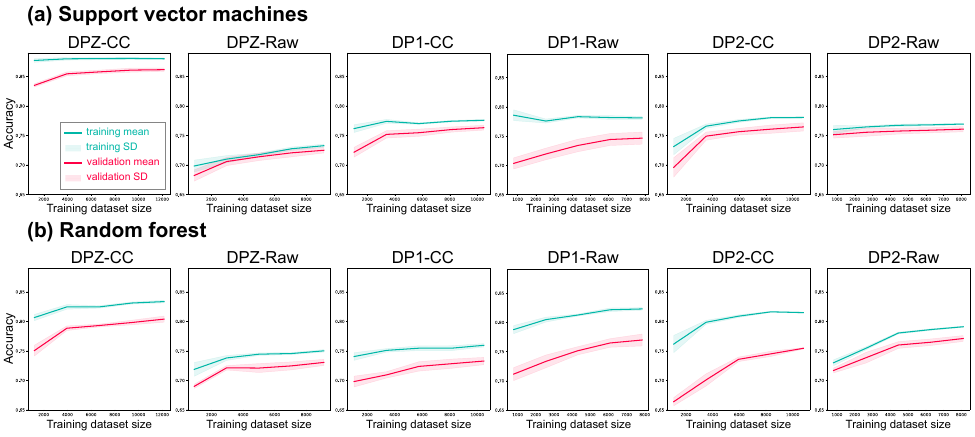}
  \caption{Learning curves of (a) SVM and (b) RF models trained on six input signals.}
  \label{fig:supp-learning}
\end{figure*}

\subsubsection{Learning curves}

Learning curves describe the accuracy of a model as a function of the size of the training set (i.e., number of training samples). \Cref{fig:supp-learning} depicts learning curves for all 12 machine learning models considered in this study obtained in cross-validation setting. The two curves for each model represent mean accuracy over $k=5$ folds of the training and validation sets as well as their standard deviation (shown as shaded regions). The training and validation sets for each of the training size were obtained by $80:20$ split. 

Note how the SVM model trained on cross-correlations of the DPZ component stands out compared to all other cases. This model has higher values of accuracy (both training and cross validation) obtained even with relatively small training sets (as low as \SI{32.5}{\percent}).

%First, we focus on the DPZ component. For each region (forest or non-forest), we use the cross-correlation signals obtained in the previous section. In the forest region, for instance, the 33 stations form 528 unique station pairs per day. For each pair, the daily cross-correlation signals are linearly stacked to produce a single cross-correlation trace using the \texttt{Stack} command in \texttt{MSNoise}. An FFT is then applied to these traces to obtain the corresponding spectra (amplitude versus frequency).

\subsection{Geometric phase analysis}

\subsubsection{State vector and geometric phase}

The geometric phase, $\Delta \eta(\omega)$, was calculated to quantify changes in the geometry of the ambient seismic wavefield between forested and non-forested regions. In this framework, the wavefield at a given frequency is represented as a high-dimensional state vector in a Hilbert space, whose components correspond to the cross-correlation signals between a reference station and multiple other stations. The geometric phase describes the angular rotation of this vector relative to a reference state and thus measures topological changes in the wavefield beyond traditional amplitude or phase quantities \cite{deymier2017sound, Luo2025Geometric}. 

The state vector, denoted $\mathbf{C}_{\omega}$, was constructed from the FFT spectra of the stacked cross-correlation signals (see Data Processing).  For a selected reference station in a given region, cross-correlations with $N$ other stations define an $N$-dimensional complex vector for each frequency, $\omega$:

%e FFT spectra of the stacked cross-correlation signals (see Data Processing) were used to construct the state vectors denoted $\mathbf{C}_{\omega}$. For a selected reference station in a given region, assembling cross-correlations between the reference station and $N$ other stations results in an $N$-dimensional complex vector for each frequency, $\omega$:

\begin{equation}
\mathbf{C}_{\omega}
= \frac{1}{\sqrt{A_1^2 + A_2^2 + A_3^2 + \cdots + A_N^2}},
\begin{pmatrix}
A_1 e^{i\phi_1} \\
A_2 e^{i\phi_2} \\
\vdots \\
A_N e^{i\phi_N}
\end{pmatrix}
\end{equation}

\noindent where $A_j$ and $\phi_j$ are the FFT magnitude and phase of the cross-correlation between the reference and $j^\text{th}$ station. The normalization ensures unit magnitude of $\mathbf{C}_{\omega}$.

%The state vector defined above depends on the selection of a reference station, we repeat the calculation for each station iteratively designated as the reference. 
This calculation was repeated for each station sequentially designated as the reference. For each reference, $N$ other stations were randomly selected from the remaining 32 stations to form the $N$-dimensional state vector, with $N=4,7,10$ explored in this study. The total number of possible combinations (e.g., $6.45\times10^7$ for $N=10$) was computationally intractable so that  400 random combinations were sampled and averaged to obtain a single state vector for each region and frequency.  Convergence tests confirmed that 400 samples were sufficient (see below).

\begin{figure}[h]
  \centering
  \includegraphics[width=\linewidth]{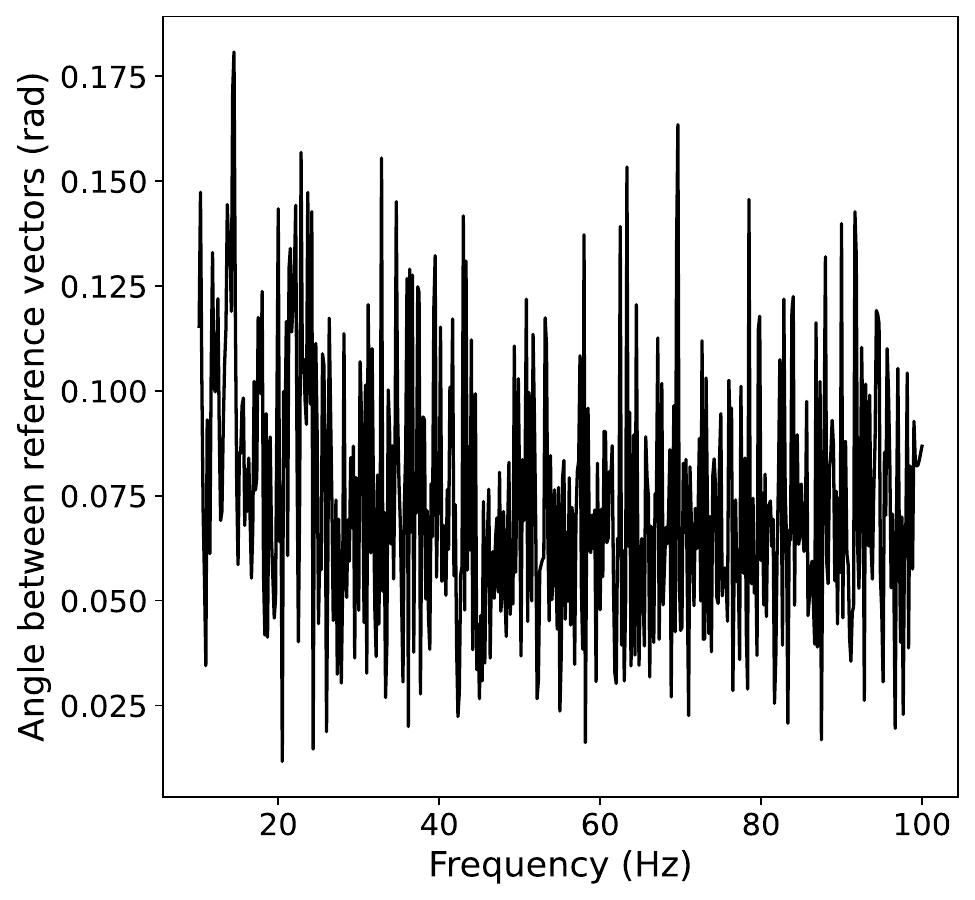}
  \caption{The angle between mean reference vectors of forest and non-forest regions. }
  \label{fig:ref_vec-diff}
\end{figure}

The geometric phase change, $\Delta\eta \in [0, \pi]$, was then calculated as the angle between $\mathbf{C}_\omega$ and its corresponding reference vector in the $N$-dimensional Hilbert space:

\begin{equation}
\Delta\eta(\omega) = \arccos[\Re(\mathbf{C}_{\omega,\text{ref}}^* \cdot \mathbf{C}_{\omega})],
\end{equation}

\noindent where $^{*}$ denotes complex conjugation and $\Re$ indicates the real part. The resulting $\Delta\eta(\omega)$ quantifies the frequency-dependent geometric perturbations of the wavefield for each region. The reference vector is defined as the average of the 400 combinations for each reference station at each frequency, providing a baseline for measuring the rotation angle in each region. This approach emphasizes the spread in the distribution of geometric phase change per region. 

Since the reference vectors differ between forested and non-forested areas, we additionally computed the angle between the reference vectors of the two regions. Specifically, for $N=10$, we first average the reference vectors across all reference stations in both regions and then compute the angle between these two vectors. The resulting angle has no distinct trend in respect to frequency and oscillates around the value of \SI{0.075}{\radian} (\Cref{fig:ref_vec-diff}). Given the small angle and low sensitivity to frequency, the distinction between forested and non-forested regions is mostly manifested in differences in the state vector dispersion and geometric phase change (Fig.\ 7 and Discussion section).

\subsubsection{Convergence test}

To make the computation tractable, we employed a subset of station combinations when constructing high-dimensional state vectors.  
The subset size must be sufficiently large such that repeated random selections yield consistent geometric phase values, $\Delta \eta$, yet small enough to remain computationally feasible. To avoid large variances in $\Delta \eta$ with deficient subsets, we assessed convergence by evaluating the variance of $\Delta \eta$ across repeated random draws. 

\begin{figure}[h]
  \centering
  \includegraphics[width=\linewidth]{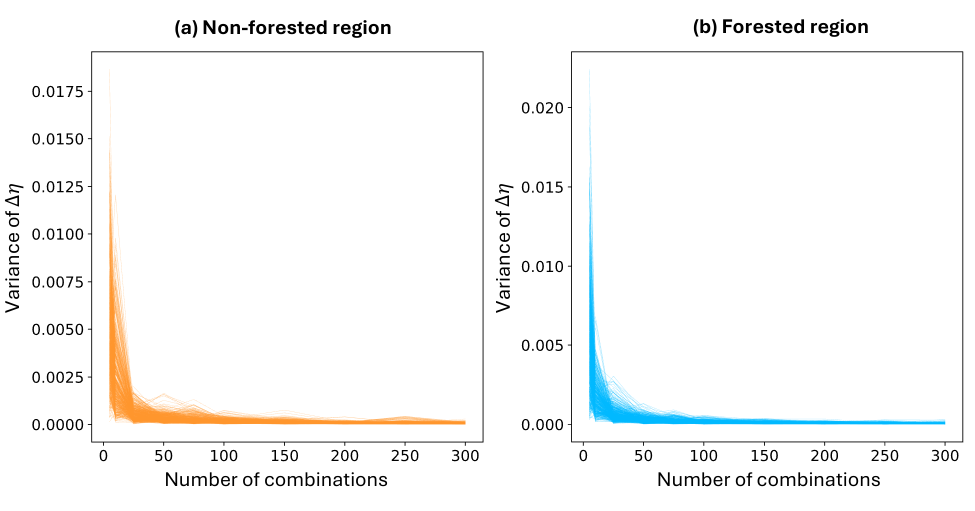}
  \caption{Convergence test for the combination set size with a fixed reference station for the (a)  non-forested (reference station \#258) and (b) forested (reference station \#83) regions. Each curve represents variance vs.\ size at a single frequency.}
  \label{fig:supp-convergence}
\end{figure}

Convergence tests were carried out with one fixed reference station and varying subset sizes (5, 10, 25, 50, 75, 100, 150, 200, 250, 300). For each size, the specified number of combinations was randomly sampled 20 times for calculation of the corresponding average values of $\Delta \eta$ at each frequency. The variance from 20 independent estimates is shown in \Cref{fig:supp-convergence}.

For both regions, the variance approaches zero once the subset size exceeds 200, which indicates converged $\Delta \eta$. To ensure robust averaging, while maintaining computational efficiency, the subset size of 400 was adopted for analysis.

\bibliography{refs.bib}{}

\end{document}